\documentclass[fleqn,twoside,twocolumn,nofootinbib]{revtex4} 
\usepackage{ujp} 
\begin{document}
\title[FLUORECENTLY LABELED BIONANOTRANSPORTERS]
{FLUORECENTLY LABELED BIONANOTRANSPORTERS\\ OF NUCLEIC ACID BASED ON
CARBON NANOTUBES}%
\author{D.S. NOVOPASHINA}
\affiliation{Institute of Chemical Biology and Fundamental Medicine of the SB RAS}
\address{8, Acad. Lavrentiev Ave., Novosibirsk 630090, Russia}
\email{danov@niboch.nsc.ru}
\author{E.K. APARTSIN}%
\affiliation{Institute of Chemical Biology and Fundamental Medicine of the SB RAS}%
\address{8, Acad. Lavrentiev Ave., Novosibirsk 630090, Russia}%
\email{danov@niboch.nsc.ru}%
\author{A.G. VENYAMINOVA}
\affiliation{Institute of Chemical Biology and Fundamental Medicine of the SB RAS}%
\address{8, Acad. Lavrentiev Ave., Novosibirsk 630090, Russia}%
\email{danov@niboch.nsc.ru}%
\udk{546.26+577.113.4/6}  \razd{\secix}

\setcounter{page}{718}%
\maketitle

\begin{abstract}
We propose an approach to the design of a new type of hybrids of
oligonucleotides with fluorescein-functionalized single-walled
carbon nanotubes. The approach is based on stacking interactions of
functionalized nanotubes with pyrene residues in conjugates of
oligonucleotides. The amino- and fluorescein-modified single-walled
carbon nanotubes are obtained, and their physico-chemical properties
are investigated. The effect of the functionalization type of carbon
nanotubes on the efficacy of the sorption of pyrene conjugates of
oligonucleotides was examined. The proposed non-covalent hybrids of
fluorescein-labeled carbon nanotubes with oli\-go\-nu\-cleotides may
be used for the intracellular transport of functional nucleic acids.
\end{abstract}

\section{Introduction}

Carbon nanotubes (CNTs) possessing chemical passivity and
compatibility with biomacromolecules and cells are investigated
actively in different fields of nanobiotechnology and biomedicine.
The wide variety of both covalent and non-covalent functionalization
methods have been described [1]. The application of CNTs as
transporters of biologically active compounds into living cells
requires the presence of reporter groups in their structure to carry
out the monitoring of transfection during both \textit{in vitro} and
\textit{in vivo} experiments. The presence of fluorophore residues
linked to the CNT surface could provide us the possibility to
observe the nanocomplexes by confocal microscopy.

\section{Experimental Part}

Commercial carboxylic acid functionalized single-walled carbon nanotubes
(SWNT-COOH, Sigma-Aldrich, 652490) were used in this work.

Functionalized nanotubes and their hybrids with oligonucleotides
were characterized by UV-Vis, fluorescent, and infrared
spectroscopies, Raman spectroscopy, thermogravimetric and elemental
analyses, transmission and scanning electron microscopies. A
fluorimeter Cary Eclipse (Varian Inc., USA), spectrophotometer
Shimadzu UV-vis-2100 (Shimadzu, Japan), electron microscopes LEO
1430 (LEO, Germany) and JEM 2010 (JEOL, Japan), confocal microscope
Cell Observer SD (Zeiss, USA), IR-spectrometer Scimitar FTS 2000
(Digilab, Australia), Raman spectrometer T64000 (Horiba Jobin Yvon,
Italy), CNT-analyzer Carlo Erba 1106 (Carlo Erba, Italy), and
thermoanalyzer TG 209 F1 (NETZSCH, Germany) were used for these
purposes.

Hexamethylenediamine and PAMAM G3.0 functionalized SWNTs (SWNT-HMDA
and SWNT-PAMAM) were prepared by analogy with [2,3].
Fluorescein-labeled carbon nanotubes, SWNT-HMDA-FITC and
SWNT-PAMAM-FITC, were obtained by analogy with [4,5].
Oligonucleotides were synthesized within the solid-phase
phosphoramidite method on an automatic synthesizer ASM-800 (Biosset,
Russia). Conjugates of oligodeoxyribonucleotides and
oligo(2$'$-O-methylribonucleotides) with pyrene residues attached to
the $5'$-phosphate group directly or via a hexa(ethyleneglycol)
phosphate linker were synthesized as described previously [6] and
used for the preparation of hybrids with SWNT. The structure of the
conjugates was proved by MALDI-TOF mass spectrometry, UV and
fluorescent spectroscopies.

For the preparation of hybrids, the functionalized SWNTs were
dispersed in a $5\times 10^{-6}$ М water solution of $5'$-pyrene
conjugate of oligonucleotide by ultrasonication during 30 min. The
concentrations of functionalized SWNT were varied within 2.5--250
$\mu $g/ml.

\begin{figure*}
\includegraphics[width=17cm]{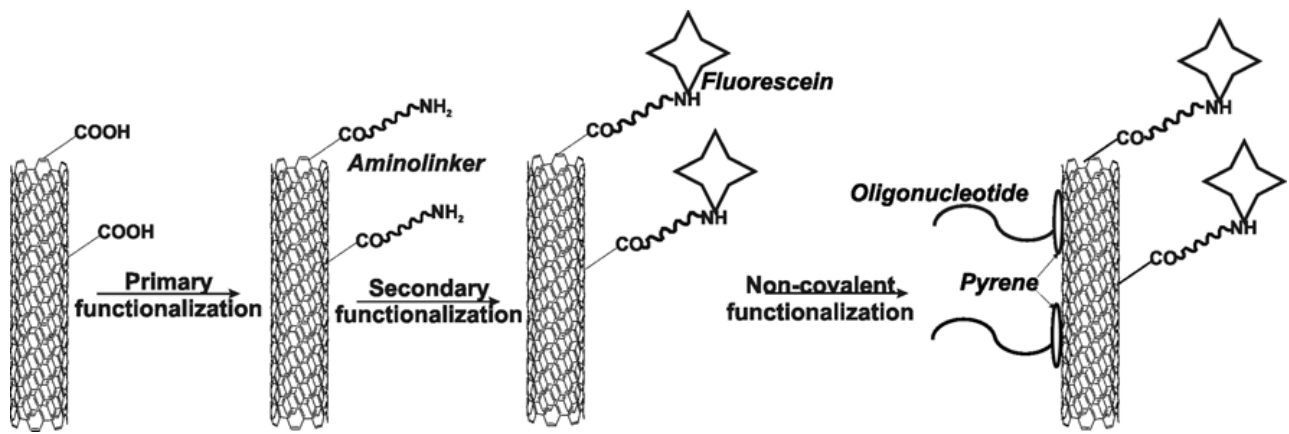}
\vskip-3mm\caption{ Strategy of SWNT multifunctionalization
}\label{fig1}\vskip3mm
\end{figure*}

\section{Results and Discussion}

Here, we propose an approach to the preparation of non-covalent hybrids of
oligonucleotides with fluorescent single-walled carbon nanotubes based on
the stacking interactions of pyrene conjugates of oligonucleotides with
the surface of nanotubes. This approach does not require severe conditions at the
stage of a modification of CNTs and is remarkable due to the relative simplicity of
the synthesis of pyrene conjugates and a higher density of CNT functionalization as
compared with other methods.

We have developed the multifunctionalization strategy including
primary  covalent SWNT functionalization by the aminolinker
introduction, secondary covalent SWNT functionalization by the
coupling of fluorescein isothiocyanate (FITC) to amino groups, and
the non-covalent attachment of $5'$-pyrene conjugates of
oligonucleotides on the SWNT surface (Fig.~1).

\begin{table}[b]
\noindent\caption{The data of thermogravimetric and elemental
analyses of functionalized SWNTs}\vskip3mm\tabcolsep2.8pt
\noindent{\footnotesize\begin{tabular}{l c c c c }
 \hline \multicolumn{1}{c}
{\rule{0pt}{9pt}Functionalized} & \multicolumn{3}{|c}{Element
content, {\%}}&
\multicolumn{1}{|c}{Weight loss upon}\\\cline{2-4}%
\multicolumn{1}{c}{SWNT}& \multicolumn{1}{|c}{C}&
\multicolumn{1}{|c}{H}& \multicolumn{1}{|c}{N}&
\multicolumn{1}{|c}{heating up }\\%
\multicolumn{1}{c}{}& \multicolumn{1}{|c}{}& \multicolumn{1}{|c}{}&
\multicolumn{1}{|c}{}& \multicolumn{1}{|c}{$400\,^{\circ}$С,\,
{\%}}\\%
\hline%
SWNT-COOH & 89.8 & 0.7 & -- & 7.7 \\%
SWNT-HMDA & 79.5 & 2.5 &4.7 & 27.8  \\%
SWNT-PAMAM & 56.2 & 5.5 &14.6 & 42.5  \\%
SWNT-HMDA-FITC & 70.7 & 3.0 &3.5 & 46  \\%
SWNT-PAMAM-FITC &63.6 &2.9 & 5.7 & 44 \\%
\hline
\end{tabular}}
\end{table}

Two types of amino-modified SWNT bearing hexamethylenediamine
residue (SWNT-HMDA) or polyamidoamine dendrimer G3.0 (SWNT-PAMAM)
were prepared. The Kaiser test [7] was employed to quantify the amount
of amino groups after the primary functionalization step (0.11--0.26
mmol/g). The fluorescein-labeled SWNTs (SWNT-HMDA-FITC and
SWNT-PAMAM-FITC) were obtained, by using these amino-modified SWNTs.
Amino- and fluorescein-modified nanotubes were characterized by
infrared spectroscopy, thermogravimetric analysis, elemental
analysis, Raman spectroscopy, and transmission and scanning electron
microscopies.

The FTIR spectra of aminomodified (SWNT-HMDA and SWNT-PAMAM) and
fluorescently labeled (SWNT-HMDA-FITC и SWNT-PAMAM-FITC) carbon nanotubes
show characteristic vibration modes corresponding to $\nu $(N--H) in amino
groups and $\nu $(C=O) and $\nu $(C--N) in amide groups. The FTIR spectrum of
SWNT-PAMAM is presented in Fig. 2 as an example.

Thermogravimetric and elemental analyses confirmed the presence of
HMDA, PAMAM, and FITC residues on the surface of functionalized SWNTs
(Table~1).

\begin{figure}
\includegraphics[width=\column]{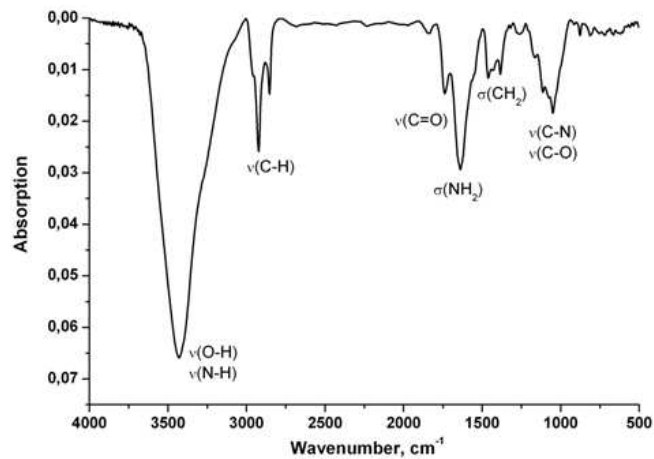}
\vskip-3mm\caption{FTIR spectrum of SWNT-PAMAM-FITC }\label{fig2}
\end{figure}

\begin{figure*}
\includegraphics[width=15cm]{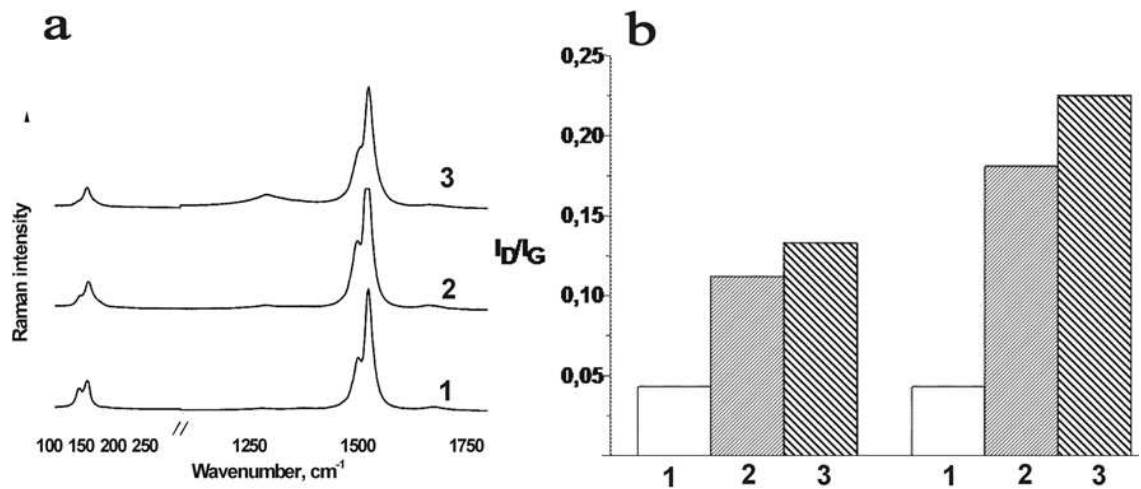}
\vskip-3mm\caption{Raman spectra of functionalized SWNTs ({\it a})
and $I_{\rm D}/I_{\rm G}$ ratios ({\it b}). {\it 1}~-- SWNT-COOH,
{\it 2}~-- SWMT-HMDA, {\it 3}~-- SWNT-HMDA-FITC
}\label{fig3}\vskip3mm
\end{figure*}

The maxima in the region of 260--290~nm (corresponded to SWNTs) and
495 nm (corresponded to fluorescein) were observed in the electronic
absorption spectra of fluorescein-labeled SWNTs. The Raman spectra
of functionalized SWNTs display the D band ($\omega _{\rm D}=1300$
cm$^{-1})$, G band ($\omega _{\rm G}=1500$--1550 cm$^{-1}),$ and the
radial breathing mode (RBM) band ($\omega _{\rm RBM}=158$
cm$^{-1}).$ The diameter was calculated according to [8] and was
equal to 1.71~nm. The ratio $I_{\rm D}/I_{\rm G}$ grows upon the
functionalization that can be considered as an indirect confirmation
of the organic molecule attachment to the SWNT surface (Fig.~3).

\begin{figure}
\includegraphics[width=\column]{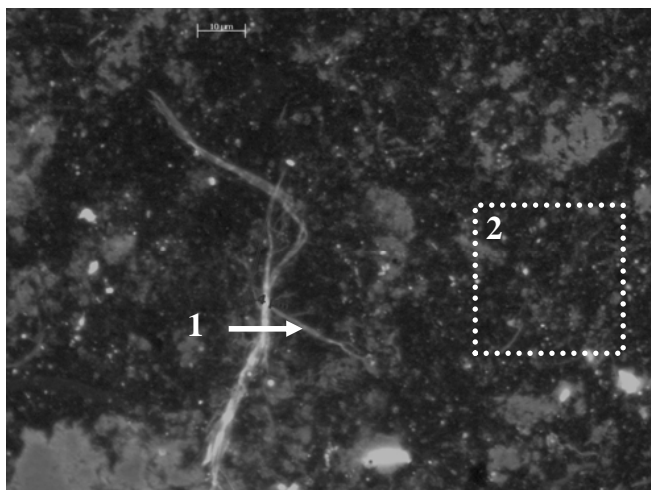}
\vskip-2mm\caption{Confocal microscopy of SWNT-HMDA-FITC. {\it 1} --
bundle of nanotubes; {\it 2} -- individual nanotubes
 }\label{fig4}
\end{figure}

The individual SWNTs with length up to 1.5 $\mu $m and their
aggregates were observed, by using SEM. The additional organic groups
were visualized as bulky structures on the ends of SWNTs. The
presence of PAMAM and PAMAM-FITC groups on the SWNT surface was
confirmed by TEM. The possibility of the visualization of fluorescently
labeled SWNTs by confocal microscopy was demonstrated (Fig. 4).

$5'$-Pyrene conjugates of oligodeoxyribonucleotides and
oligo($2'$-O-methylribonucleotides) (Fig. 5) were purified by PAAG
electrophoresis and immobilized on functionalized SWNTs.

The fluorescent properties of the obtained hybrids were investigated. The increasing
amount of SWNTs in a solution reduced the fluorescence intensity of pyrene groups
because of the fluorescence quenching upon the interaction
with nanotubes [9]. We used this phenomenon to elaborate the method of
quantitative estimation of the amount of oligonucleotides fixed on the SWNT
surface. This method may be applied for the concentration of nanotubes in
a solution up to 50 mg/ml.

The presence of bulky functional groups on the nanotube surface may
affect the rate and the completeness of the sorption of pyrene
residues. The isotherms ($25^{\circ}$C) of the absorption of
$5'$-pyrene modified oligonucleotides on the nanotubes surface are
plotted in Fig. 6.

It is shown that the rates of oligonucleotide sorption are
slightly different in the cases of modified and unmodified SWNTs.
The complete sorption of oligonucleotide (90-95{\%}) was achieved when
the concentration of SWNTs was about 50 $\mu $g/ml irrespective of the
type of nanotube functionalization. The presence of a hexaethyleneglycol linker
between the pyrene residue and an oligonucleotide decreases the oligonucleotide
sorption efficiency, if the concentration of SWNTs is less than 50
$\mu $g/ml. The capacity of a carbon nanotube was amounted to be about 100
$\mu $mol/g.

\begin{figure*}
\includegraphics[width=17cm]{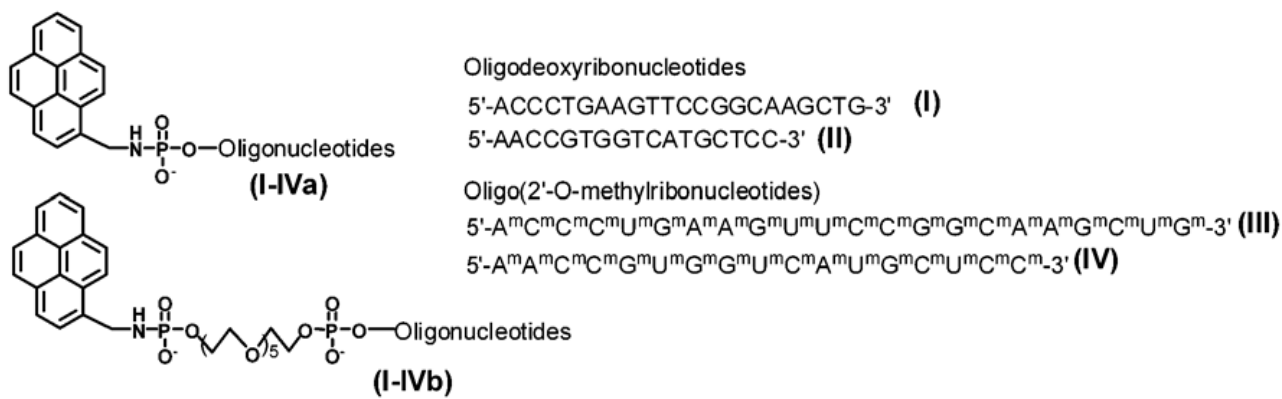}
\vskip-3mm\caption{$5'$-Pyrene-modified oligonucleotides used for
the preparation of hybrids }\label{fig5}\vskip3mm
\end{figure*}

\begin{figure}
\includegraphics[width=\column]{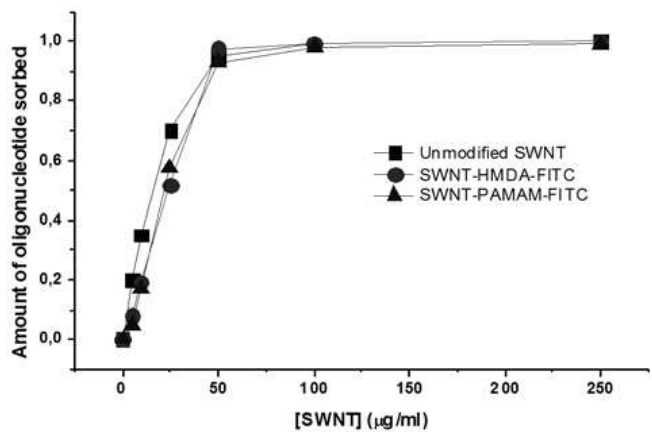}
\vskip-3mm\caption{ Isotherms of the $5'$-pyrene conjugate
(\textbf{Ia}) absorption on the surface of functionalized nanotubes.
Conditions: 25 $^{\circ}$C, oligonucleotide concentration $5\times
10^{-6}$ М }\label{fig6}
\end{figure}

The data on non-covalent hybrids of oligonucleotides with
functionalized SWNTs were obtained, by using high-resolution electron
microscopy. TEM images demonstrated the simultaneous presence of
a functional group (FITC, PAMAM, PAMAM-FITC) and an oligonucleotide in
the structure of hybrids. Oligonucleotides were visualized as folded
nanosized structures on the SWNT surface (Fig. 7).

\begin{figure}
\includegraphics[width=\column]{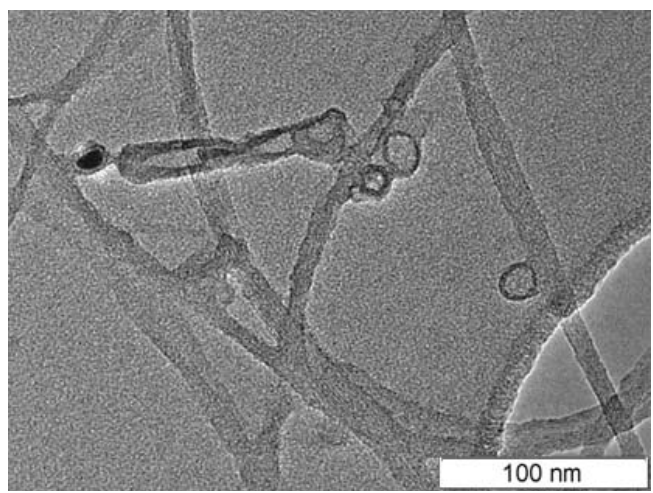}
\vskip-3mm\caption{TEM-image of the non-covalent hybrid of
$5'$-pyrene conjugate (\textbf{IIa}) with SWNT-HMDA-FITC
}\label{fig7}
\end{figure}

The investigation of the desorption of $5'$-pyrene conjugates of
oligonucleotides from the CNT surface was performed. Pyrene
conjugates of oligonucleotides were displaced from the CNT surface
by methylene blue that caused the precipitation of carbon nanotubes.
The increase of the desorbed $5'$-pyrene conjugate of
oligonucleotides in supernatant upon a rise of the methylene blue
concentration was observed by the gel electrophoresis assay. These
data confirmed the non-covalent nature of hybrids.

\begin{figure}
\includegraphics[width=\column]{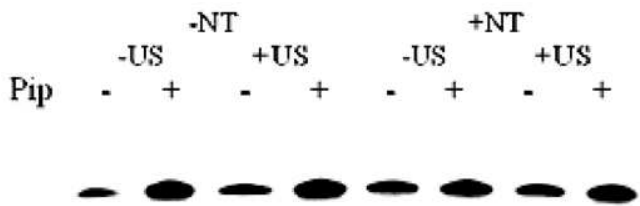}
\vskip-3mm\caption{Stability of oligonucleotide \textbf{(Ia) }during
preparation of their hybrids with functionalized SWNTs (NT) and
ultrasonication (US). Pip -- piperidine treatment }\label{fig8}
\end{figure}

The stability of oligonucleotides during the preparation of hybrids
was studied. The gel electrophoresis assay displayed no strand
breaks in oligonucleotides after the ultrasonication in the presence
of SWNT-COOH (Fig. 8). The absence of apurinic/apyrimidinic (AP)
sites in oligonucleotides attached to SWNT-COOH was demonstrated by
the piperidine treatment. The reversed phase HPLC analysis of
hydrolyzates of the same oligonucleotides did not detect any
products of nucleoside oxidation. These data proved the stability of
oligonucleotides during the formation of hybrids.

\section{Conclusions}

The results obtained demonstrate the feasibility of the proposed
approach for the design of hybrids of oligonucleotides with CNTs.
Such non-covalent hybrids may be applied to making the
bionanotransporters of functional nucleic acids (siRNA,
NA-enzymes, aptamers, {\it etc}.)

\vskip3mm The authors thank the collaborators from the Scientific and
Educational Centre of NSU ``Nanosystems and modern materials''
and Cooperative centre ``Nanostuctures'', Dr.~V.I.~Zaikovskiy for
performing the physical analysis of modified nanotubes, and
Dr.~V.V.~Koval for the registration of MALDI-TOF mass spectra.

This work was supported by the RFBR grant 11-04-01014-a, FTP
``Scientific and scientific-educational personnel of the innovative
Russia'' (grant P1334), FASIE grant, and the RAS Presidium program of
basic research No. 27 (project~62).

\rezume {%
 ФЛУОРЕСЦЕНТНО МІЧЕНІ \\ БІОНАНОТРАНСПОРТЕРИ НУКЛЕЇНОВИХ \\ КИСЛОТ НА
 ОСНОВІ ВУГЛЕЦЕВИХ НАНОТРУБОК} {Д.С. Новопашина, Є.К. Апарцін, А.Г.
 Веньямінова} {У даній роботі ми пропонуємо новий підхід до створення
 гібридів олігонуклеотидів з флуоресцентно міченими одностінними
 вуглецевими нанотрубками. Використаний нами підхід заснований на
 стекинг-взаємодії залишків пірену в складі піренільних
 кон'югатів олігонуклеотидів з поверхнею функціоналізованих
 нанотрубок. Було отримано аміно- і флуоресцеїн-модифіковані
 одностінні вуглецеві нанотрубки і вивчено їх фізико-хімічні
 властивості. Досліджено ефект впливу типу функціоналізації вуглецевих
 нанотрубок на ефективність сорбції піренільних кон'югатів
 олігонуклеотидів. Запропоновані в роботі нековалентні гібриди
 флуоресцентно мічених вуглецевих нанотрубок з олігонуклеотидами в
 перспективі можуть бути використані як транспортери функціональних
 нуклеїнових кислот у клітини.}

\end{document}